# THE GENERAL LAW PRINCIPLES
# FOR PROTECTION THE PERSONAL DATA
# AND THEIR IMPORTANCE

Jonatas S. de Souza[1, 2], Jair M, Abe[1], Luiz A. de Lima[1, 2] and
Nilson A. de Souza[2, 3]

[1]Graduate Program in Production Engineering - Paulista University, São Paulo,
Brazil
[2]National Association of Data Privacy Professionals - ANPPD - Scientific
Committee, São Paulo, Brazil
[3]São Paulo State Supplementary Pension Foundation– PREVCOM,
São Paulo, Brazil

## ABSTRACT

*Rapid technological change and globalization have created new challenges when it comes to the protection and processing of personal data. In 2018, Brazil presented a new law that has the proposal to inform how personal data should be collected and treated, to guarantee the security and integrity of the data holder. The purpose of this paper is to emphasize the principles of the General Law on Personal Data Protection, informing real cases of leakage of personal data and thus obtaining an understanding of the importance of gains that meet the interests of Internet users on the subject and its benefits to the entire Brazilian society.*

## KEYWORDS

*Data, Law General, Personal, Protection, Regulation, Legal.*

## 1. INTRODUCTION

The concern about the protection of people's data has grown over the years, but only after the approval of the Brazilian Law that received the name of Marco Civil da Internet, established by Law No. 12,965, 2014 [1]. In Brazil, a new law has recently been sanctioned and it is generating a lot of discussion in several areas. The General Law on Personal Data Protection, Law No. 13.709 [2] of 14[th] August 2018, gives the Brazilian population rights and guarantees on how organizations will have to adapt to the collection and processing of personal data, whether by physical or digital means.

Discussing data protection in Brazil has become a challenging task. The state of the crisis provoked by COVID-19 (coronavirus) had a severe impact on companies, which began to adopt measures to make their workforce compatible with the demands existing during social isolation, and the adoption of measures to minimize the risk of the disease spreading among their workforce.

The use of Virtual Private Network - VPN and practices such as BYOD (Bring Your Own Device) have become common to incorporate daily life. There was also an exponential growth of e-commerce, home office, webinars, virtual meetings, and numerous activities that started to occur entirely through the Internet.

In the same proportion, the risks associated with the improper use of personal data, data leaks, improper access by third parties, theft of data kept by corporate servers, creation of fake profiles, fake news, among other practices frequently reported were multiplied. The objective of the paper is to present important aspects such as the principles and fundamentals of Brazilian law and to present some real cases on data leaks.

The paper is composed of sections, in Section 2 presents the Theoretical Reference that will address the history of data protection in Brazil and the European Regulation, in Section 3 describes the Principles of Brazilian Law demonstrating the similarity with the European Regulation, in Section 4 the Importance of Brazilian Law showing the fundamentals of the Law and emphasizes the importance of consent of the data holder, in Section 5 the Results and Discussions with real cases of data leaks and countries that already have some legislation on protection of personal data, in Section 6 are the Conclusions bringing the final considerations obtained.

## 2. THEORETICAL REFERENCE

### 2.1. Personal Data Protection in Brazil

In Brazil, the legislation is based on the positivist model of law, adopted by Lusitanian, German and Italian schools that privilege the written law, this reflects in the delay of the implementation of the legislative process (figure 1), which begins with the initial idea, passes through the creation of the bill, then through the bicameral approval and then the presidential sanction, to finally come into force with coercive force.

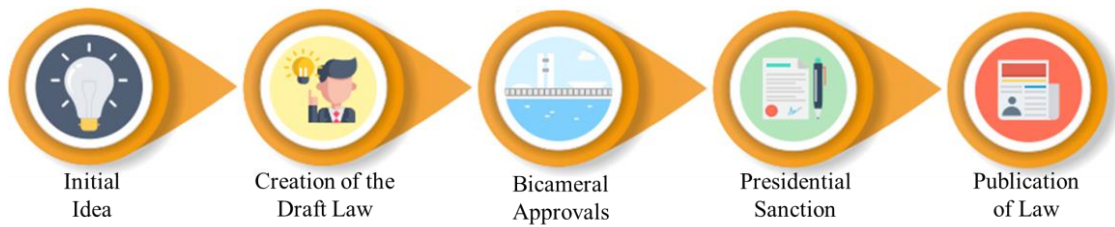

Initial          Creation of the       Bicameral       Presidential        Publication
Idea             Draft Law             Approvals        Sanction            of Law

Fig. 1: Law Creation Process.

The first Brazilian initiative on personal data protection was in Article 5 of the 1988 Federal Constitution [3].

Art. 5 All are equal the law, without distinction of any nature, guaranteeing Brazilians and foreigners residing in the country the inviolability of the right to life, freedom, equality, security, and property, under the following terms [3]:

X - The intimacy, privacy, honor, and image of persons are inviolable, and the right to compensation for material or non-material damage resulting from their violation is guaranteed [3].

XII - The secrecy of correspondence and telegraphic communications, data, and telephone communications shall be inviolable, except in the latter case by judicial order, in the cases and the manner established by law for criminal investigation or criminal proceedings [3].

Law No. 9.296, of July 24th, 1996 [4] deals with the interception of telephone communications and regulates clauses XII, Art. 5 of the Federal Constitution.

On September 11th, 1990 [5] Law No. 8.078, known as the Consumer Code (CDC), was enacted, bringing in its Article 43 the guarantee of access to the holder's data, demanding clarity and objectivity of the information and the possibility for the consumer to demand the correction of his registration data [5].

Art. 43 The consumer, without prejudice to the provisions of Art. 86, shall have access to the information existing in registers, files, records, personal data, and consumption filed about him, as well as to their respective sources [5].

Paragraph 1. Consumer registrations and data must be objective, clear, truthful, and in easy-to-understand language, and may not contain negative information for a period longer than five years [5].

Paragraph 2. The opening of the registration, file, record, personal, and consumption data shall be communicated in writing to the consumer when not requested by him [5].

Paragraph 3. The consumer, whenever he finds any inaccuracy in his data and registrations, may demand their immediate correction, and the archivist shall, within five working days, communicate the change to the eventual recipients of the incorrect information [5].

Paragraph 4. Databases and registers relating to consumers, credit protection services, and the similar are considered public entities [5].

Paragraph 5. Once the statute of limitations on the collection of consumer debts has been consummated, the respective Credit Protection Systems shall not provide any information that may prevent or hinder new access to credit with suppliers [5].

Paragraph 6. All information referred to in the caption of this article must be made available in accessible formats, including for persons with a disability, at the request of the consumer [5].
Even bringing some progress on personal data protection, the CDC was still limited in its scope on the subject, which means that the protection would exist in the relationship between supplier and consumer within the scope of the legal concepts established in Articles 2 and Article 3 [5] of the CDC.

On April 23rd, 2014, Law No. 12,965, now known as Marco Civil da Internet [1], was approved, establishing principles, guarantees, rights, and duties for the use of the Internet in Brazil, and has the guarantee of privacy and protection of personal data, and will only make such data available through a court order. In Art. 7, clauses I, II and III, and clauses VII, VIII, IX, and X, deal with the rights of the holders of personal data [1].

Art. 7 Access to the Internet is essential to the exercise of citizenship, and the user has assured the following rights [1]:

I - Inviolability of intimacy and privacy, their protection and compensation for material or moral damage resulting from their violation [1].

II - Inviolability and secrecy of the flow of your communications over the Internet, except by court order, in the form of the law [1].

III - Inviolability and secrecy of your stored private communications, except by court order [1].

VII – Do not provide third parties with your data, including connection records, and access to internet applications, except by free, express and informed consent or in the cases provided by law [1].

VIII - Clear and complete information about the collection, use, storage, treatment and protection of your data, which may only be used for purposes that: a) justify their collection; b) are not prohibited by law, and c) are specified in service contracts or terms of use of internet applications [1].

IX - Express consent on the collection, use, storage, and processing of personal data, which shall occur in a manner detached from the other contractual clauses [1].

X - Definitive exclusion of personal data that you have provided to a certain internet application, at your request, at the end of the agreement between the parties, except for the cases of mandatory storage of records provided for in this Law [1].

The Civil Framework of the Internet also includes aspects of the responsibility for the protection of personal data by access providers and in operations carried out through the Internet, providing for some sanctions, described in Articles 10, 11, and 12 [1].

On August 4th, 2018, Law No. 13,709, called the General Law on Personal Data Protection [6], was approved, providing for the processing of personal data, whether digital or not, to protect the fundamental rights of freedom and privacy and the development personal personality of the individual in society.

## 2.2. General Law on Personal Data Protection

The General Law on Personal Data Protection - LGPD, Law No. 13.709 of 14th August 2018, which would come into force in August 2020, has been postponed by Provisional Measure No. 959/2020 [6] extending the vacatio legis [7] and postponed to 3 May 2021 [8]. The LGPD purpose is to provide guidelines on how personal data will be collected and processed, and to ensure the security and integrity of the data holder, whether digital or not. On 10th July 2018, Project Law 53/2018 - PLC [9] was approved by the plenary of the Federal Senate and was sanctioned on 14th August 2018 by the 37th President of Brazil [2]. Article 1 of the LGPD states that it is prepared to protect the processing of personal data to protect the rights of freedom, privacy, and personality development of the individual. Moreover, it applies to any individual or legal entity that carries out-processing operations such as collection, production, reception, classification, processing, among other activities by physical or digital means in Brazilian territory, or abroad if it is using personal data of individuals living in Brazil.

## 2.3. General Data Protection Regulation

The General Data Protection Regulation 2016/679 [10] – GDPR, of the European Parliament and of the Council of European Union – EU, of 27th April 2016, is a Regulation that is on the protection of individuals about the processing of personal data and the free movement of such data and that repeals Directive 95/46/EC [11], EU companies had two years to comply with the regulation by the date of 28th May 2018. The Regulation applies to all activities involving the processing of personal data using full or partial consent, as well as to the processing of personal data by non-automated means.

# 3. THE PRINCIPLES OF LGPD

For a better understanding of LGPD [2], it is necessary to know the legal bases (principles) that should be observed for any type of data processing activities, the Law is composed of ten principles that are listed in Art 6. A GDPR [10] [12] is also guided by principles [13], which are set out in Article 5, which form the basis for the EU Regulation, and these principles should be linked to data processing.

## 3.1. Principle of Purpose

The LGPD [2], the purpose for which the data will be done must be very specific, explicit, and informed to the holder of the personal data that will be processed [2]. In GDPR, the Purpose Limitation principle [13], the data must be collected for specific, legitimate, and explicit purposes, and may not be processed for other unspecified purposes [10].

## 3.2. Principle of Adequacy

The LGPD [2], is the formality with the holder of the personal data to process personal data [2]. In GDPR, the Storage Limitation principle [13], data may be stored in a database until the end of the data processing and must be informed to the data owner, and after the end of the processing, the data must be deleted from the database. It is also linked to the principle of Bidding [13] that the company that will process the data must comply with the Regulation and with the data holder. [10].

## 3.3. Principle of Necessity

In the LGPD [2], the amount of data for data processing is only relevant, proportional, and not excessive [2]. In the European Regulation, the Data Minimization principle [13], data should be collected following its purpose and only data that are necessary for the processing [10].

## 3.4. Principle of Free Access

The LGPD [2], guarantees that the data holder will have free access to the data in its entirety at any time, and this principle is linked to the GDPR Transparency principle. In the European Regulation there is a right which is described in Article 17 [10], which is called Right to Erasure [10] or Right of Forgetfulness, which gives the "right to be forgotten" to the data holder of the database concerning the purpose of the processing, after the data holder has requested to delete the data, the officer shall delete the data relating to the data holder's request [10].

## 3.5. The Principle of Data Quality

The LGPD [2], guarantees the data owner clarity, accuracy, and relevance and updates the data according to the needs of the data treatment [2]. The Accuracy principle of GDPR [13] that data should always be updated and correct thus maintaining the quality of the data that will be processed and incorrect data will be rectified or deleted [10].

## 3.6. Principle of Transparency

The LGPD [2], ensures that the data owner will have access to all necessary information clearly, accurately, and easy access to data processing [2]. The Transparency Principle of GDPR is divided into three words, Lawfulness, Fairness, and Transparency [10] [13]. The Lawfulness or Bidding is concerned, data controllers should comply with the Regulation, on Fairness or Loyalty, it is stated that processing should take place fairly with the consent of the data owner, and on Transparency, the data controller will allow him to have access to all information of the data processing [10].

### 3.7. Principle of Security

The LGPD [2], will use techniques for the protection of personal data from unauthorized access or accidental or illicit situations of alteration, destruction, loss, dissemination, and communication [2]. The principle that about security in GDPR is the Integrity principle and Confidentiality [13], the data must be stored securely, guaranteeing the data integrity, and adopting methods of protection against unauthorized processing, loss, accidental damage, destruction, or unauthorized access [10].

### 3.8. Principle of Prevention

It will use methods to prevent data processing damaging [2] [10].

### 3.9. Principle of Non-Discrimination

Data may not be processed for discrimination, illicit or abusive purposes [2].

### 3.10. Principle of Accountability and Reporting

In Brazilian Law [2], it is up to the treatment agent to prove the purpose and which effective methods have been adopted, and he must be able to prove compliance with and enforcement of personal data protection rules, including the effectiveness of these methods [2]. In the European Regulation, the Accountability principle [13], which is the full responsibility of the data processing agent, guarantees the length of the purpose of the processing and has evidence of the necessity of the processing [10].

## 4. THE IMPORTANCE OF LGPD

The LGPD [2], sanctioned in Brazil, was inspired by GDPR of the EU [10] and contains many similarities in its respective principles. In its Art. 2 they show the foundations (figure 2), that served as a basis for the development of the Law [2]:

Art. 2 The discipline of personal data protection based on the according to fundamentals [2]:

I - Respect for privacy [2].
II - Informative Self-determination [2].
III - Freedom of Expression, Information, Communication, and Opinion [2].
IV - The Inviolability of Intimacy, Honour, and Image [2].
V - Economic and Technological Development and Innovation [2].
VI - Free Enterprise, Free Competition, and Consumer Protection [2].
VII - Human Rights, Free Development of Personality, Dignity, and the exercise of citizenship by natural persons [2].

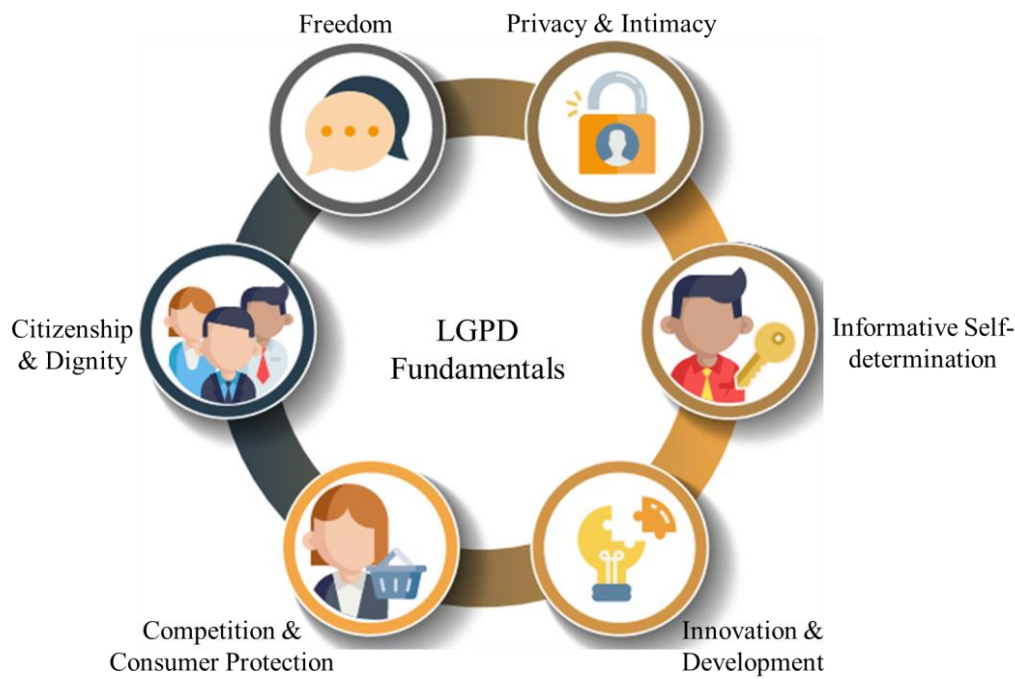

Fig. 2: LGPD Fundamentals [14].

One of the most important points that makes data processing possible is to have the consent of the data holder, according to Article 7, clauses 1 [2], and in Article 8 [2] it is reinforced that the authorization must be in writing or by other means of manifestation of the owner will, and it is stated from Paragraph 2 [2] that in case the authorization is in writing it must be highlighted in the contractual clauses. In the case of processing sensitive personal data, Article 11, clauses 1 [2], states that with the consent of the holder, and Article 14, Paragraph 1 [2], which states must have the consent of parents or legal guardians concerning the processing of personal data of children and adolescents.

In GDPR [10] it is also explicit that for any activities that require a data processing must have the consent of the data holder, in Article 6, Paragraph 1, clauses A [10], it says that data processing will be lawful upon the consent of the data holder for specified purposes previously informed to the data holder, in Article 7 [10] which sets out the conditions applicable to consent, it says that the data processing agent must prove that the data holder has agreed to the specified purposes. As regards the processing of data on children, Article 8 of the European Regulation [10] requires the person legally responsible for the child under the age 16 to consent to the processing. The State may be responsible for giving consent if the child is under the age of 13 and has no family members to answer for him or her.

Without a reference law for the use of personal data, the possibility of abuse in the collection and use of personal data is increased, as well as the encouragement of several other non-specialized bodies to issue their opinions regarding the use of data, which causes great confusion. This is the case, for example, of inspections and inspections by the Public Prosecutor's Office and consumer protection agencies, the issuance of opinions by regulatory agencies, or even judicial decisions based on various sparse legal provisions [1] [5] that seek to define parameters for the processing of personal data.

In the graph (figure 3), shows the level of interest in Internet users searches on the terms LGPD and GDPR over a twelve-month period, where the term LGPD represents the blue line and the term GDPR represents the red line, on the horizontal axis represents the time and on the vertical axis represents the level of search made on the terms, these levels are represented by the numbers 0 (very low), 25 (low), 50 (average), 75 (high) and 100 (very high). This simple analysis shows that the red line had several peaks in some periods, this because the GDPR since 2016 [10] is approved and had an adequacy period of two years and the level of interest is between average and very high, the blue line has had small peaks, this because the LGPD is a new subject in Brazil and this makes the level of interest is between very low and average. In general, users looking for LGPD and GDPR terms are professionals in the juridical environment or Information Technology.

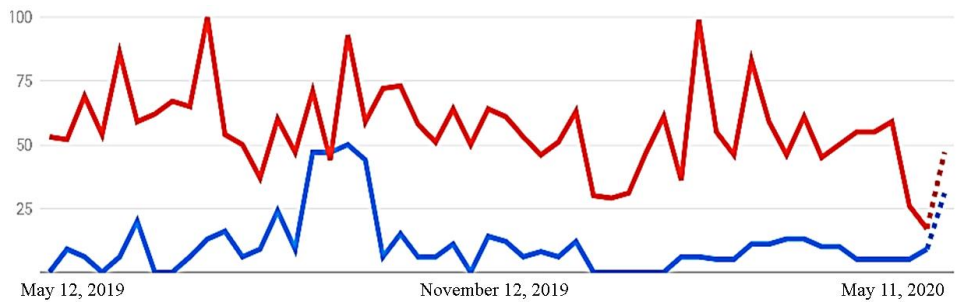

Fig. 3: Level of interest in the terms LGPD and GPDR [25].

## 5. DISCUSSIONS

It should be noted that both the Brazilian Law and the European Regulation, they provide a guide for data processing and what procedures companies should take to comply with the law if these principles are not followed these companies will be at serious legal risk. An example of non-compliance with the law was the Cambridge Analytica scandal [15], which misused data from 87 million Facebook users (figure 4), manipulated the data without the consent of the data holders, to help win Donald Trump's US presidential campaign, and for the British to vote to leave the European Union, both in 2016 [15], Facebook was asked about data security.

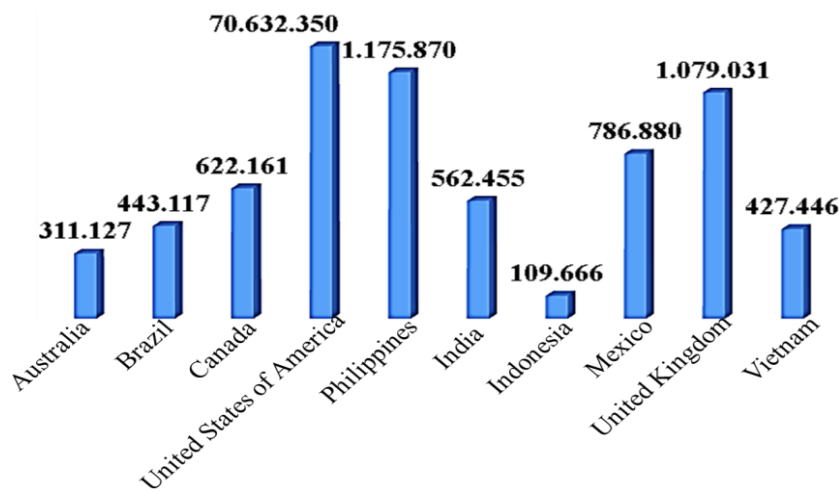

Fig. 4: Number of Facebook users who may have had their data used improperly with Cambridge Analytica [16].

In Brazil, there have been several cases of data loss, such as the case of the Netshoes website, according to the Coordinator of the Commission for Personal Data Protection, Prosecutor Frederico Meinberg, "this is one of the largest security incidents recorded in Brazil" [17], which because of the data leak could put the integrity of 1,999,704 users at risk if the leaked data fell into the wrong hands.

The impact that data leaks go far beyond the financial losses, the exposure of each citizens' information can be irreversible damage that becomes impossible to measure the size of the loss. Without an information security policy, it can cause serious problems such as the invasion of vital systems to steal tax returns, data, making illegal financial transfers, interrupting the strategic operations of a company, or the government.

Another case about data leakage was written by Liliane Nakagawa and published on the website Olhar Digital [18], which displays the news about the banking institution, specifically the Bank of Brazil Provident Fund [17]. According to Nakagawa, data leak that reaches 153 thousand clients - official number of registered in the BB Previdencia platform, according to Bank of Brazil. The source, who identified the security gap, stated that through the private pension system, aimed at companies and public agencies, it is possible to have access to all personal data of participants and, from breaking, editing and registering beneficiaries, all in the name of the registered person himself [18].

After this news, several headlines were reporting the incident, the Exame magazine published on its website, "BB Previdencia website leak exposes data of 153 thousand clients" [19], the newspaper, O Estado de S. Paulo, published on its website, "Security sheet on BB Previdencia website exposes client data" [20] (figure 5).

Fig. 5: Personal Data Empty from the BBTurPrev Plan Withdrawal Page [18].

For these leaks not to occur, companies must have a Data Protection Officer – DPO [2] [10], where the primary function is to ensure that the organization processes the personal data of their employees, their customers, their suppliers or any other individuals securely and reliably according to the data protection rules of law [10].

An LGPD will give the right to protection of the personal data of the respective holders and will give guidelines to the companies on how the treatment should be done. Brazil will be adapting to GPDR and will move the job market for data protection specialists. However, Brazil [21] already has a law for the creation of a supervisory body to verify whether companies comply with the LGPD, but directors have not yet been appointed to the National Data Protection Authority - ANPD and the National Council for Personal Data Protection and Privacy [8]. The European body responsible for supervising undertakings on whether they comply with the European Regulation is the European Data Protection Supervisor - EDPS, an independent supervisory authority established according to EU Regulation 2018/1725, and its task is to ensure that the fundamental rights and freedoms of individuals - in particular their privacy - are respected when EU institutions and bodies process personal data. In the world, there are already some countries [22] outside the EU that have a regulation regarding data protection. On the European Commission's website, it informs countries that are at an appropriate level to the Regulation, the European Commission has recognized Andorra, Argentina [23], Canada (trade organizations), Faeroe Islands, Guernsey, Israel, Isle of Man, Jersey, New Zealand, Switzerland, Uruguay [23] and the United States of America (limited to the Privacy Shield framework) as providing adequate protection [24].

## 6. CONCLUSIONS

Through the Internet Civil Framework, which establishes rights and duties, guarantees and principles for the use of the Internet in Brazil, it does not guarantee data protection and privacy in a well-structured, complete and comprehensive manner, nor is a general regulation on the protection of personal data, and its provisions on data protection not protective in nature.

Some of the challenges identified for implementing the Law in Brazil are legal adjustments and appropriate training, a complete action plan for companies to comply with LGPD, specialized implementation of personal data governance processes, information security technologies, educating Brazilian society about this Law by showing the rights and duties of citizens.

Therefore, there will still be a lot of debates and discussions about LGPD and whether it will adhere to GDPR, and how Brazil will behave with the law when it becomes effective.

### ACKNOWLEDGMENTS

This study was financed in part by the Coordenação de Aperfeiçoamento de Pessoal de Nível Superior - Brasil (CAPES) -Finance Code 001.

## AUTHORS

**Jonatas Santos de Souza** has a degree in Information Systems from the Facudade Impacta de Tecnologia – FIT (2016). Has experience in Computer Science, with emphasis on Information Systems, working mainly on the following topics: Artificial Intelligence, Paraconsistent Analysis Network, Paraconsistent Logic, Industry 4.0, and Artificial Neurons. He holds a Post-Graduation in Management and Governance of Information Technology by Anhanguera (2019), and a Master's Degree in Production Engineering by Universidade Paulista - UNIP, being a CAPES scholarship holder. Member of the Scientific Committee of the National Association of Data Protection Professionals- ANPPD.

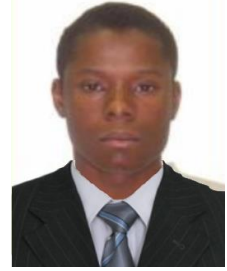

**Jair Minoro Abe** received B.A. and MSc in Pure Mathematics - University of Sao Paulo, Brazil. I also received the Doctor Degree and Livre-Docente title from the same University. He is currently the coordinator of the Logic Area of Institute of Advanced Studies - University of Sao Paulo Brazil and Full Professor at Paulista University - Brazil. His research interest topics include Paraconsistent Annotated Logics and AI, ANN in Biomedicine, and Automation, among others. He is a Senior Member of IEEE. Leader of the Research Group "Paraconsistent Logic and Artificial Intelligence" cataloged at CNPq Fellowship,

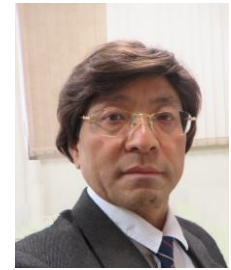

**Luiz Antonio de Lima** Doctor of Science student in Production Engineering Universidade Paulista, Master's degree in Production Engineering in the area of Artificial Intelligence Applied to Software Paraconsistent Measurement Software, Post-Undergraduate Degree in EAD, University Professor, General Coordinator of IT Course and Campus Assistant: 2008-2009. Speaker and Event Organizer: SENAED; NETLOG; WICS; WINFORMA. IT Consultant and/or roles: IT Director, Commercial Director, Project Manager, with clients: WCI-MahlerTI, WCI-Ericsson, WCI-Frema, WCI-Brasil Brokers, WCI-Gt1, The WCI-Consoft-IPESP, WCI-Consoft-SPPREV, WCI-Consoft-PMSP, WCI-Eversystems-Rede Globo / BankBoston, and WCI- Consoft-BankBoston: using best practices in the market: ITIL, COBIT, SIX SIGMA - Black Belt, CMMi, PMBOK, SCRUM, APF, APT, UCP.

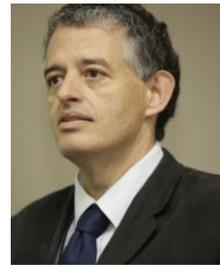

**Nilson Amado de Souza** has a degree in Pedagogy from the University of North Paraná (2010). Has experience in Science and Technology, with emphasis on Information Technology POST-GRADUATE - Teaching for Higher Education FALC - Faculty Aldeia de Carapicuíba POST-GRADUATE - Specialization in Business Intelligence Faculty Impacta de Tecnologia - FIT, ITIL Expert certification; ISO 20000 certification; Privacy Data certification; DPO Privacy and Data Protection Foundation certification; Privacy and Data Protection Practitioner Privacy and Data Protection Essentials ISO-27001

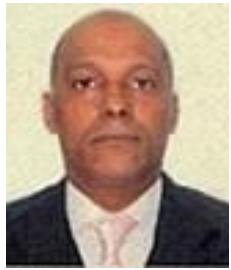